В.С. Усатюк (аспирант)

# ЗАДАЧИ ТЕОРИИ РЕШЕТОК И ИХ ВЗАИМНЫЕ РЕДУКЦИИ

Братск, Братский государственный университет

Задачи теории решеток лежат в основе целого класса криптографических примитивов и протоколов «постквантовой криптографии»:

− ассиметричных системах шифрования: Айтая-Дворка(Ajtai-Dwork), Реджева(Regev), Джентри (Gentry), NTRU [1];

− криптографических хеш-функциях: Айтая ([2],[3]), LASH [4], SWIFFT [5], SWIFFTX [6];

− протоколах цифровой подписи: Гольдвассера-Гольдштейна-Халеви (GGH)[7], $NTRUS_{IGN}$ [8], Джентри-Пейкерта-Вейкантанатана[9], Миссиансио-Вадхена[10], Миссиансио-Любашевского[11];

− протоколах защищенного обмена данными и идентификационных схемах: Джентри-Пейкерта-Вейкантанатана (IBE) [9], Пейкерта-Вейкантанатана [12], Пейкерта-Вейкантанатана-Вотерса(OT) [13], Любашевского [14].

Поэтому изучение и уточнение свойств задач теории решеток это одна из основных целей, как при построении, так и при криптоанализе примитивов и протоколов на основе задач теории решеток. Важность этого направления исследований обусловлена так же наличием у задач теории решеток (при некоторых параметрах) свойств криптостойкости к алгоритмам, выполняемым на квантовых компьютерах. Роль последнего обстоятельства, в свете недавнего открытия коллективом ученных под руководством Джереми О'Брайена механизма создания фотонных квантовых компьютеров посредством традиционной литографии СБИС, резко возрастает ([15], [16]).

Решетка – дискретная аддитивная подгруппа, заданная на множестве $R^n$, то есть решетку $L$ можно представить как множество целочисленных линейных комбинаций $L(b_1,...,b_2) = \{\sum_{i=1}^{n} x_i b_i : x_1,...,x_n \in Z\}$, $n$ – линейно независи-мых базисных векторов $\{\bar{b}_1,...,\bar{b}_n\} \subset R^m$ в $m$-мерном Евклидовом, где $m$ и $n$, размерность и ранг решетки соответственно (рис. 1). У решетки может быть множество базисов, $L = \sum_{i=1}^{n} \bar{a}_i \cdot Z$, (рис. 2). На рисунках 3, 4 показаны фундаментальные параллелепипеды образованные базисами. Площади (объемы в многомерном случае) фундаментальных параллелепипедов образованных всевозможными базисами одной решетки $L$, $\det(L)$ будут равны, $\det L$ - инвариант решетки. Под кратчайшим вектором решетки будем понимать вектор, длина которого $\lambda(L) = \min_{x,y \in L, x \neq y} \|x-y\| = \min_{x \in L, x \neq 0} \|x\|$ (рис. 5). Тогда многомерным обобщением этого понятия для решетки размерностью $n$, $L^n$ будет $i$–й последовательный минимум $\lambda_i(L)$, под которым понимают наименьший радиус окружности(шара) содержащий $i$–линейно независимых векторов $\lambda_i^p(L) = r$, $r \in R : \exists v_i \in L, \max_i \|v_i\| \leq r$, где $v_i$-это линейно независимые вектора (рис. 6). Первый последовательный минимум в решетке $\lambda_1^p(L)$ соответствует длине кратчайшего вектора в решетке. Нормы определяют в контексте каждой конкретной задачи, однако обычно используются нормы $\ell_1 = \max_j \sum_i |a_{ij}|$, $\ell_2$ - Евклида или $\ell_\infty = \|(x_1, x_2,...,x_n)\|_\infty = \max_{i=1,2,...,n} |x_i|$ - Чебышева.

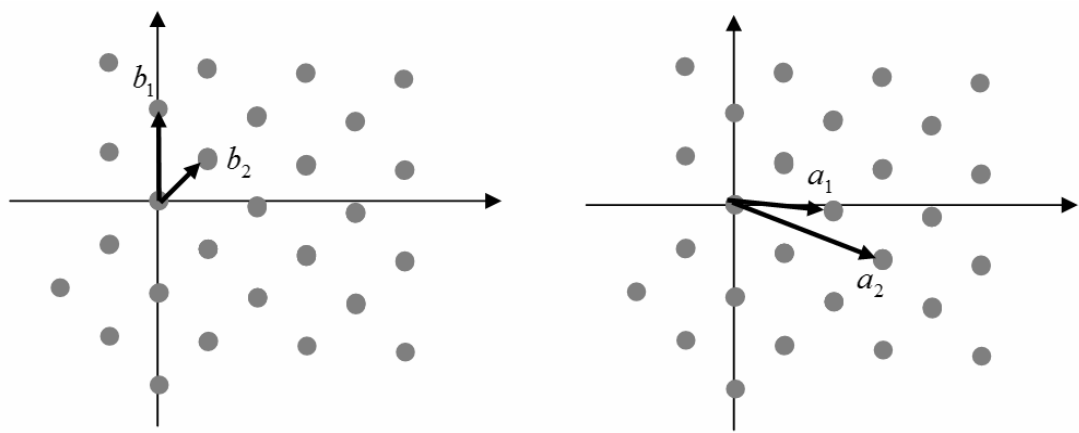

Рис. 1. Решетка с базисом $\{\bar{b}_1, \bar{b}_2\} \in B$   Рис. 2. Решетка с базисом $\{\bar{a}_1, \bar{a}_2\} \in B$

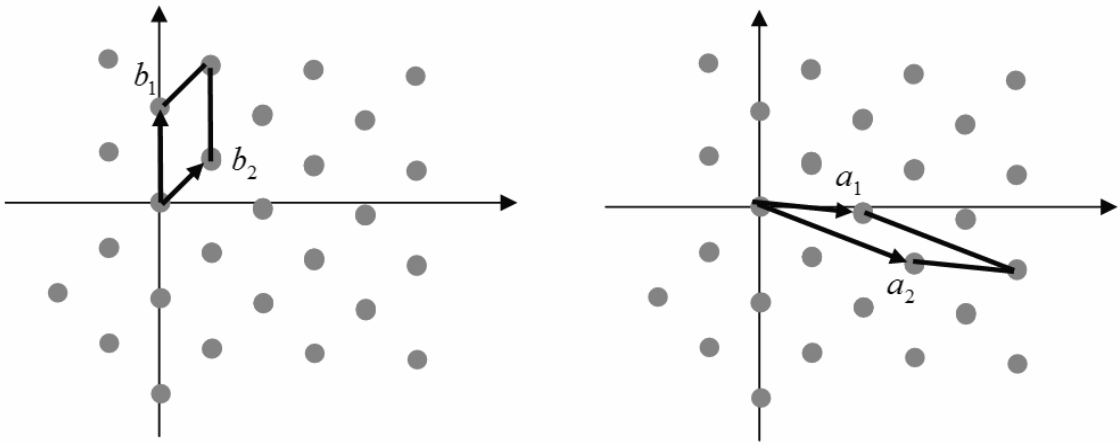

Рис. 3, 4. Фундаментальные параллелепипеды, образованные базисами

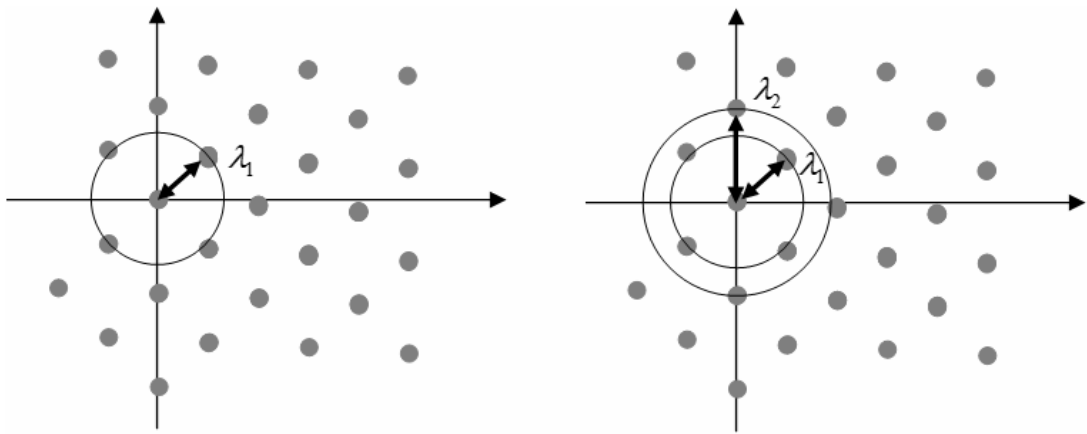

Рис. 5. Кратчайший вектор    Рис. 6. Кратчайший базис решетки $L$ в $R^2$

Идеальная решетка – это решетка со свойствами идеала на некотором кольце чисел, то есть результат сложения и умножения векторов в такой решетке так же принадлежит ей самой. Пусть задано кольцо многочленов с целочисленным коэффициентами $R = Z[x]/f(x)$ взятых по модулю $f(x)$, где $f(x)$ - унитарный приведенный многочлен степени $n$. Так как заданное отношение изоморфно над $Z^n$ и аддитивные группы и идеалы кольца являются подгруппами в $R$, полученная конструкция представляет собой решетку. Такой класс решеток принято называть идеальными решетками.

Таким образом, познакомившись с основными определениями теории решеток, перечислим задачи, применяемые при построении и оценке сложности в рассматриваемых нами системах шифрования[17]:

1. По базису решетки найти кратчайший ненулевой вектор (shortest vector problem, SVP; поиск кратчайшего вектора), (рис. 7);

2. По базису решетки $B \in Z^{m \times n}$ и вещественному $\gamma > 0$, найти ненулевой вектор $\bar{b} \in BZ^n \setminus \{0\} : \|\bar{b}\|_p \leq \gamma \cdot \lambda^p{}_1(L)$ с p-нормой ($\gamma$-approximation shortest vector problem, $\text{SVP}^p_\gamma$; приближенный поиск кратчайшего вектора), (рис. 8);

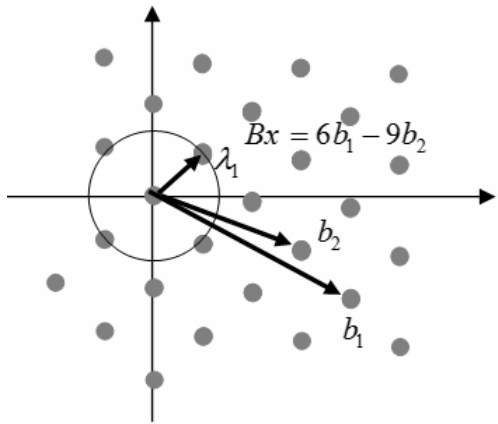 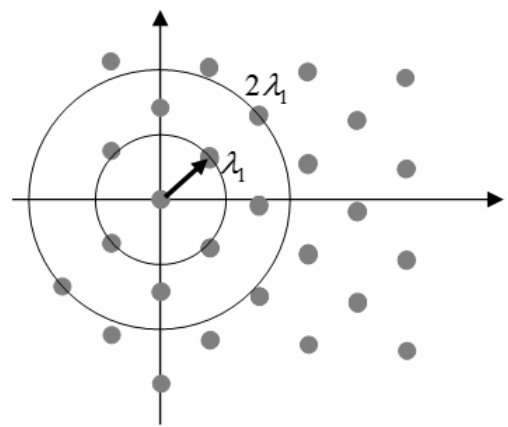

Рис. 7. Пример SVP-задачи в $R^2$     Рис. 8. Пример $SVP_\gamma$-задачи в $R^2$

3. По базису решетки $B$ и заданному вектору $\bar{j} \notin L(B)$, найти ближайший вектор $\bar{b} \in L(B)$ (Closes vector problem, CVP; поиск ближайшего вектора), является неоднородным(гетерогенным) вариантом SVP-задачи), (рис. 9);

4. По базису решетки $B \in Z^{m \times n}$, вещественному $\gamma > 0$ и заданному вектору $\bar{j} \in LR^n$, найти ненулевой вектор $\bar{b} \in BZ^n : \|\bar{j} - \bar{b}\|_p \leq \gamma \lambda_1^p(L)$ с p-нормой ($\gamma$-approximate closes vector problem, $\text{CVP}^p_\gamma$);

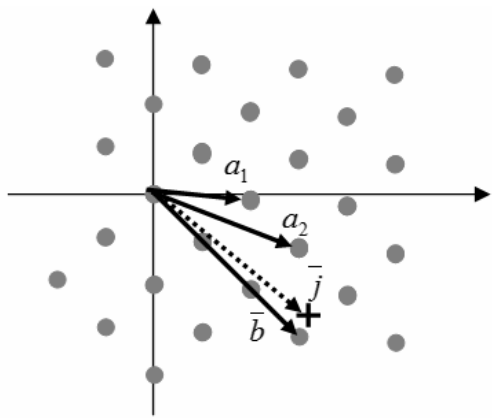 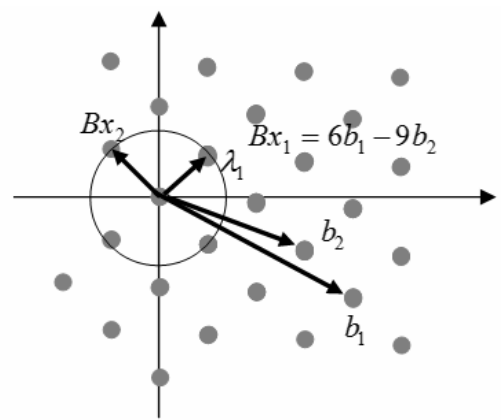

Рис. 9. Пример CVP-задачи в $R^2$     Рис. 10. Пример SIVP-задачи в $R^2$

5. Пусть дана n-мерная решетка L. Найти линейно независимые вектора $\bar{b}_1,...,\bar{b}_n \in L$, для которых $\max_{i=1}^{n}\|\bar{b}_i\|_p \leq \gamma \lambda^p_n(L)$, где $\lambda^p_n(L)$ - $i$ – й последовательный минимум в решетке с $p$-нормой ($\gamma$-approximate shortest independent vector problem, $SIVP_\gamma^p(n,\gamma)$; приближенный поиск кратчайших линейно независимых векторов), (рис. 10);

6. Пусть дана n-мерная решетка L. Найти вектор $\bar{u} \in L \setminus \{0\} : \|u\|_p \leq \gamma \lambda_1^p(L)$, где $\lambda^p_1(L)$, это длина кратчайшего вектора в решетке с p-нормой и $\bar{u}$ - кратчайший $\gamma$ - уникальный вектор, т.е. $\forall w \in L : \lambda_1^p(L) \leq \|\bar{w}\|_p \leq \gamma \lambda_1^p(L)$  $\bar{w} = z\bar{u}$ для некоторых $z \in Z$ ($\gamma$-approximate unique shortest vector problem, $uSVP_\gamma^p(n,\gamma)$; поиск уникального кратчайшего вектора), (рис. 11);

7. Пусть дан базис $B$, $q$-нарной(модулярной) $m$-мерной решетки $L_q^{mxn}$, т.е. решетка L, для которой принадлежность вектора к решетки L определяется: $L(B) = \{B^T s \bmod q \subseteq Z^m, s \in Z^n\}$, $q$ – простое число. На решетке равномерно распределен шум $e$ (обычно с моментом ожидания равным 0 и дисперсией $\sqrt{q}$), q задан некоторым многочленом, $\bar{s} \in Z_q^n$ - некоторый исходный вектор без шума, известно значение $(B\bar{s} + \bar{e})$. Найти исходную точку в решетке (исключить шум) по некоторому множеству известных $(B\bar{s}_i + \bar{e}_i)$. Задача обучения с ошибками (Learning with errors, $LWE$) является обобщением задачи обучения контроля целостности (четности) данных с шумами, (Рис. 12);

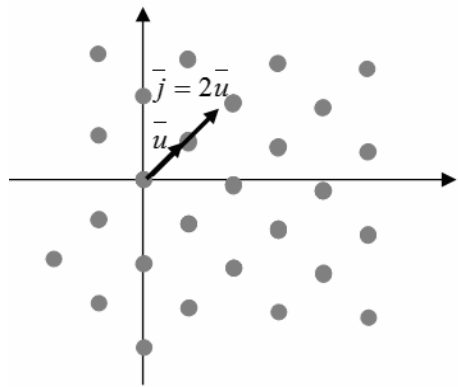 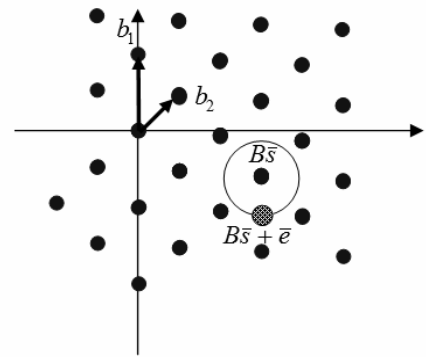

Рис. 11. Пример uSVP-задачи     Рис. 12. Пример LWE-задачи

8. Пусть дана $m$-мерная модулярная решетка $L_q^\perp(B) = \{\bar{x} \in Z^m : B\bar{x} \equiv \bar{0}Z^n \bmod q\}$ образованная базисом $B \in Z_q^{n \times m}$, взятым случайно из равномерного распределения над $Z_q^{n \times m}$. Найти вектор кратчайший вектор $\bar{v} \in L_q^\perp(A) : \|\bar{v}\|_p \leq \beta$ с $p$-нормой (Short integer solution problem, $\mathrm{SIS}_\beta^p(n,m)$; задача поиска вектора по норме в модулярной решетке);

9. Пусть дана $m$-мерная модулярная решетка $L_q^\perp(B) = \{\bar{x} \in Z^m : B\bar{x} \equiv \bar{0}Z^n \bmod q\}$ образованная базисом $B \in Z_q^{n \times m}$ и вектор $\bar{y} \in Z^n$, взятые случайно из равномерного распределения над $Z_q^{n \times m}$, $Z^n$ соответственно. Найти вектор $\bar{v} \in \{\bar{x} \in Z^m : A\bar{x} \equiv \bar{y} \bmod q\} : \|\bar{v}\|_p \leq \beta$. (Short integer solution problem, $\mathrm{ISIS}_\beta^p(n,m)$; гетерогенная задача поиска вектора по норме в модулярной решетке);

10. По базису решетки $B \in Z^{m \times n}$ и вещественному $\gamma \geq 1$, найти ненулевой вектор $\bar{b} \in BZ^n \setminus \{0\} : \|\bar{b}\|_p \leq \gamma \cdot \det(L(B))^{1/n}$ с $p$-нормой ($\gamma$-approximate Hermite shortest vector problem, $\mathrm{hermitSVP}_\gamma^p$; задача приближенного поиска кратчайшего вектора по Эрмиту);

11. Пусть дана $n$-мерная решетка $L^n \subseteq R^k$. Найти базис $B : \forall B' \in \{B \in Q^{n \times k} : L = L(B)\}, \max_{i=1}^n \{\|\bar{b}_i\|_p\} \leq \gamma \max_{i=1}^n \{\|\bar{b}_i'\|_p\}$ ($\gamma$-approximate shortest basis problem, $\mathrm{SBP}_\gamma^p(n)$; приближенный поиск кратчайшего базиса);

12. Пусть $n$-мерная решетка $L^n$, заданна некоторая $p$-норма. Найти длину $l^{(p)} \in R : l^{(p)} \leq \lambda_1^{(p)}(L) \leq \gamma l^{(p)}$, где $\lambda_1^{(p)}(L)$ - длина кратчайшего вектора или первый последовательный минимум в решетке $L$ ($\gamma$-approximate shortest length problem, $\mathrm{SLP}_\gamma^p(n)$; задача приближенного поиска длины кратчайшего вектора в решетке);

13. По базису решетки $B \in Z^{m \times n}, m \geq n$ и радиусу $r \in R$ ответить на вопрос: «да» - если все точки решетки можно покрыть радиусом $r$, $r \geq p(L(B))$;

«нет» - если $\gamma \cdot r < \rho(L(B))$ ($\gamma$-approximate covering radius problem, $\mathrm{CRP}_\gamma^p(n,r)$; приближенная задача покрытия решетки радиусами);

14. По базису решетки $B \in Z^{m \times n}$, вещественному $\alpha \geq 1$ и вектору $\bar{u} : \exists \bar{t} \in L(B), \|\bar{u} - \bar{t}\|_p < \alpha \cdot \lambda_1^p(L)$, найти вектор $\bar{v} \in L(B) : \|\bar{u} - \bar{v}\|_p = \min$ ($\alpha$-approximate bounded distance decoding, $\mathrm{BDD}_\alpha^p(u)$; приближенная задача о декодирование с ограниченным расстоянием);

15. По базису $B \in Z^{m \times n}$, вектору решетки $v \in BZ^n$ и положительным действительным числам $d, \gamma > 0$ ответить на вопрос: «да» - если $\min\{\|v\|_p : v \in BZ^n \setminus 0\} \leq d$; «нет» - если $\min\{\|v\|_p : v \in BZ^n \setminus 0\} > \gamma d$ (Decisional shortest vector problem, $\mathrm{GapSVP}_\gamma^p$; Булева задача о поиске кратчайшего вектора);

16. По базису $B \in Z^{m \times n}$, вектору решетки $t \in BZ^n$ и положительным действительным числам $d, \gamma > 0$ ответить на вопрос: «да» - если $\min\{\|t - v\|_p : v \in BZ^n\} \leq d$; «нет» - если $\min\{\|t - v\|_p : v \in BZ^n\} > \gamma d$ (Decisional closest vector problem, $\mathrm{GapSVP}_\gamma^p$; Булева задача поиска вектора близкого к вектору в решетке);

17. Пусть дан идеал $I \in Z(x)/f(x)$. Найти многочлен $g(x) \in I \setminus \{0\} : \|g \bmod f(x)\|_p \leq \gamma \lambda_1^p(I)$. (Approximate ideal shortest vector problem/Shortest polynomial problem, $\mathrm{Ideal\text{-}SVP}_\gamma^p(f)$; приближенная задача поиска кратчайшего вектора в идеальной решетке/задача поиска кратчайшего полинома);

18. Пусть даны $m$-многочленов $g_1(x), ..., g_m(x)$ выбранных случайно из равномерного распределения заданного на $Z_q(x)/f(x)$ и $n$ – степень многочлена $f(x)$. Найти целые $e_1, ..., e_m \in Z(x) : \sum_{i \leq m} e_i g_i = 0(\bmod q), \|e\|_p \leq \beta$, где вектор $e$ - получается путем конкатенацией(объединением) коэффициентов при всех $e_i$ (Ideal small integer solution problem, $\mathrm{Ideal\text{-}SIS}_\beta^p(g,n,m)$; задача поиска вектора по норме в идеальной решетке).

Рассмотрев задачи теории решеток, приведем ключевую теорему, лежащую в основе шифрования на основе задач теории решеток, **теорема Айтая** [2]: Определим для любого натурального $n \in N$, класс случайных решеток $L^n$ порождаемых с полиномиальной временной сложностью. Предположим, что существует полиномиальный по временной сложности алгоритм A такой, что для любой случайно выбранной решетки $L \subset L^n$, найдет нетривиальный вектор $\bar{v}$, длина которого не превосходит $n$. Значит, существует вероятностный полиномиальный по временной сложности алгоритм B, который для любой решетки $L \subset R^n$ и некоторых констант $c_0, c_1, c_2$ с высокой вероятностью способен решить любую из следующих задач:

- $SVP_\gamma$ -задачу с точностью $\gamma = n^{c_2}$;
- $SIVP_\gamma$ -задачу с точностью $\gamma = n^{c_0}$;
- $SBP_\gamma$ - задачу с точностью $\gamma = n^{c_1}$.

Этой теоремой Айтай установил связь между сложностью в худшем и среднем случаях вышеперечисленных задач, а так же продемонстрировал механизм создания односторонних функций. Цай и Неруркар в 1997, снизили значение констант: $c_0 > 3$, $c_1 > 3.5, c_2 > 4$ ([18]). Даниель Миссиансио и Одед Реджев в 2004 г. показали, что $c_2 = 1$ ([19]). Гольдштейн, Гольдвассер и Халеви исследуя односторонние функции Айтая, доказали наличие у некоторых классов решеток более сильных свойств - свойств криптографических хеш-функций [3]. Именно наличие связи между сложностью в худшем и среднем случаях позволяет построить некоторую систему шифрования, на основе одной из задач теории решеток взятой из известного случайного распределения, тем самым получив строгую оценку сложности расшифровки этой системы (криптостойкости) в худшем случае. Следует так же отметить, что большинство криптографических примитивов и протоколов (RSA, ECC и прочие) не обладают данным свойством, а основаны на сложности в среднем, что значительно затрудняет, как выбор параметров шифрования, так и строгое доказательство сложности их расшифровки.

Каждая из задач теории решеток, в зависимости от параметров (например, точности решения), принадлежит к некоторому классу временной/емкостной сложности для детерминированной и недетерминированной машин Тьюринга[20]. При анализе задач, внимание концентрируется именно на временной сложности, как ключевом параметре защищенности (собственно емкостная сложность заведомо не превышает временную). Задачи допускают взаимные редукции, что упрощает их исследование, например:

$SBP_\gamma^p(n) \leq_p SIVP_\gamma^p(n)$, ([2]); $uSVP_\gamma \leq BDD_{1/\gamma} \leq uSVP_{2/\gamma}$ ([21]); $GapCVP_\gamma^p \equiv_p CVP_\gamma^p$, $SVP_\gamma^2 \leq_p CVP_\gamma^2$ [10].

Покажем на примере редукцию $SVP^p$ - и $CVP^p$ -задач([22], [23]).

От $CVP^p$ к $SVP^p$: Пусть дан почти ортогональный базис образующий n-мерную решетку $L(B): B = \{\bar{b}_1,...,\bar{b}_n\}$. Определим решетку $L'(B'): B' = \{2\bar{b}_1,...,\bar{b}_n\}$ и решим $CVP^p$-задачу для вектора $\bar{b}_1$ и решетки $L'(B')$, получив вектор $\bar{v} \in L'$. Вычислим новый вектор $\bar{s}_1 = \bar{v} - \bar{b}_1$. Выполним предыдущие действия для векторов $\bar{b}_2,...,\bar{b}_n$, получив вектора $\bar{s}_2,...,\bar{s}_n$. Найдем кратчайший из векторов $\bar{s}_1,...,\bar{s}_n$.

От $SVP^p$ к $CVP^p$: Пусть дан базис образующий n-мерную решетку $L_0^n(B'): B' = \{\bar{b}_1,...,\bar{b}_n\}, B \in Z^n$ и некоторая точка $y \notin L^n(B')$. Решим первую часть $SVP^p$-задачи, выполнив ортогонализацию базиса, как необходимое условие поиска кратчайшего вектора. Вычислять длинны векторов, а так же выбирать среди них кратчайший вектор, в данном случае нет необходимости. Получим решетку $L^n(B) \equiv L_0^n(B')$. Найдем вектор $\bar{a} \in R^n$, решив линейную систему уравнений $B\bar{a} = y$. Округлив координаты полученного вектора до целых значений, получим искомый вектор $\bar{z} \in Z^n$.

Сложность $CVP^p$-задачи превосходит сложность $SVP^p$-задачи, так как необходимым элементом решения $CVP$-задачи является наиболее ресурсоемкая часть задачи поиска кратчайшего вектора, а именно приведение базиса решетки к ортогональному виду. Сложность вычисления длин векторов будет зависеть от нормы ($O(n\log n)$ для норм Евклида $\ell_2$, [24]), а сложность нахождения минимального элемента будет $O(n)$ в худшем случае. Причем от качества базиса будет зависеть погрешность $CVP^p$-алгоритма, для предложенного алгоритма $\varepsilon = \|B\overline{a} - B\overline{z}\| \leq \frac{1}{2}\left\|\sum_i b_i\right\|$. Например для решеток $L_1(B_1) \equiv L_2(B_2), |\det(B_1)| = |\det(B_2)|$ образованных базисами $B_1 = \{(1,0),(0,1)\}$, $B_2 = \{(100,1),(99,1)\}$, погрешность будет $\varepsilon_1 = 1.4, \varepsilon_2 \approx 199$ соответственно.

Задачи теории решеток предоставляют широкие возможности реализации криптографических протоколов и примитивов, а так же затрагивают множество фундаментальных вопросов, начиная от задач линейного программирования и многомерной упаковки тел до обобщенных теоретико-числовых проблем. Построение криптографических систем на основе задач теории решеток невозможно без дальнейшего исследования вышеперечисленных задач и алгоритмов их решения.

**Список литературы.**